\documentclass{PoS}
\usepackage{epstopdf,url}
\usepackage{lineno}
\title{Sensitivity for tau neutrinos at PeV energies and beyond with the MAGIC telescopes}

\ShortTitle{Search  for tau neutrinos at PeV energies and beyond with the MAGIC telescopes}

\abstract{The MAGIC telescopes, located at the Roque de los Muchachos Observatory (2200 a.s.l.) in the Canary Island of La Palma, are placed on the top of a mountain, from where a window of visibility of about 5 deg in zenith and 80 deg in azimuth is open in the direction of the surrounding ocean. This permits to search for a signature of particle showers induced by earth-skimming cosmic tau neutrinos in the PeV to EeV energy range arising from the ocean.
We have studied the response of MAGIC to such events, employing Monte Carlo simulations of upward-going tau neutrino showers. The analysis of the shower images shows that air showers induced by tau neutrinos can be discriminated from the hadronic background coming from a similar direction. We have calculated the point source acceptance and the expected event rates, assuming an incoming tau neutrino flux consistent with IceCube measurements, and for a sample of generic neutrino fluxes from photo-hadronic interactions in AGNs. The analysis of about 30 hours of data taken toward the sea leads to a point source sensitivity for tau neutrinos at the level of the down-going point source analysis of the Pierre Auger Observatory.\
      }

\author{D. G\'ora$^{1}$, M. Manganaro$^{2,3}$, E. Bernardini$^{4,5}$, M. Doro$^{6}$, M. Will$^{2,3}$, S. Lombardi$^{7}$, J. Rico$^{8}$, D. Sobczynska$^{9}$, \speaker{J. Palacio$^{8}$}, for the MAGIC Collaboration \thanks{https://magic.mpp.mpg.de/acknowledgements\_19\_05\_2017.html}\\
        E-mail: \email{Dariusz.Gora@ifj.edu.pl}}

\author{\\                                                                                                                                                                               
        $^1$Institute of Nuclear Physics Polish Academy of Sciences, PL-31342 Krakow, Poland\\
        $^2$Inst. de Astrofisica de Canarias, E-38200 La Laguna, Tenerife, Spain\\  
        $^3$Universidad de La Laguna, Dpto. Astrof'isica, E-38206 La Laguna, Tenerife, Spain\\                                                                                                                                                               $^4$Deutsches Elektronen-Synchrotron (DESY), D-15738 Zeuthen, Germany\\
        $^5$Humboldt University of Berlin, Institut f\''ur Physik Newtonstr. 15, 12489 Berlin Germany\\
        $^6$Universita di Padova and INFN, I-35131 Padova, Italy\\
        $^7$INAF National Institute for Astrophysics, I-00136 Rome,Italy\\
        $^8$Institut de Fisica d'Altes Energies (IFAE), The Barcelona Institute of Science and Technology, Campus UAB, 08193 Bellaterra (Barcelona), Spain\\
        $^9$University of  L\'odz, PL-90236 Lodz, Poland\\
 }

\FullConference{35th International Cosmic Ray Conference -- ICRC217-\\
		10-20 July, 2017\\
		Bexco, Busan, Korea}

\begin{document}
\section{Introduction}
\vspace{-0.5 cm}
 A conventional approach for the detection of neutrinos with energies in the PeV range is based on detectors which use large volumes of ice (IceCube) or water (ANTARES).  They sample Cherenkov light from  muons produced by muon neutrinos, or from electron and tau lepton induced  showers initiated by the charged current interactions of electron and tau neutrinos. An alternative technique is based on the observation of
upward going extensive air showers  produced by the leptons originating from neutrino interactions below the surface of the Earth, 
the  so-called earth-skimming method~\cite{fargion,bertou}.  Neutrino induced showers could be detected using a variety of  methods: surface particle detector arrays and air fluorescence telescopes, like in the Pierre Auger Observatory  and the Telescope Array\footnote{\url{https://www.auger.org/} and  \url{http://www.telescopearray.org/}}, or radio detectors like ANITA\footnote{\url{http://www.phys.hawaii.edu/~anita/new/html/science.html} }. 
Another possibility is to use the technique of Imaging Atmospheric-Cherenkov
Telescopes (IACTs) which is widely used in contemporary gamma-ray astronomy. An IACT system  capable to detect the neutrino-induced  signature should look to the ground, e.g. the side of a mountain or  the sea surface~\cite{fargion,Asaoka:2012em,gora:2016}. 

In this paper we study the possibility to use the  MAGIC (Major Atmospheric Gamma Imaging Cherenkov) telescopes to search  for air showers induced by tau neutrinos 
($\tau$-induced showers) in the PeV-EeV energy range. MAGIC is a system of  two  IACTs located at the Roque de los Muchachos Observatory (28.8$^{\circ}$ N, 17.9$^{\circ}$ W), in the Canary Island of La Palma (Spain). They are placed 85 m apart, each with a primary mirror of 17 m diameter. The MAGIC telescopes, with a field of view (FOV) of 3.5$^{\circ}$, are able to detect cosmic $\gamma$-rays in the energy range 50 GeV - 50 TeV~\cite{magicperformance}. 

In order to  use MAGIC for tau neutrino searches, the telescopes need to be pointed in the direction of the tau neutrinos escaping first from the Earth crust and then from the ocean, i.e. at the horizon or a few degrees below. More precisely, at a zenith angle of 91.5$^{\circ}$ the surface of the sea is $165$ km away, thus the telescopes can monitor a large volume in  their FOV resulting  in a space angle area (defined as the intersection of the telescopes  FOV  and  the sea surface) of about a few km$^2$. 
 In~\cite{upgoing_magic}, the effective area for up-going tau neutrino observations with the MAGIC telescopes was calculated analytically and found to reach  $\sim 10^3$ m$^2$ (at 100 TeV) and  10$^5$ m$^2$ (at 1 EeV) for an observation angle of about 
1.5$^{\circ}$ below the horizon, rapidly diminishing with higher inclination. However, the  sensitivity for diffuse neutrinos was found to be  very poor because of the limited FOV, the observation time, and the low expected neutrino flux.
\begin{figure*}[t!]
 \centering
 \noindent

\includegraphics [width=0.27\textwidth]{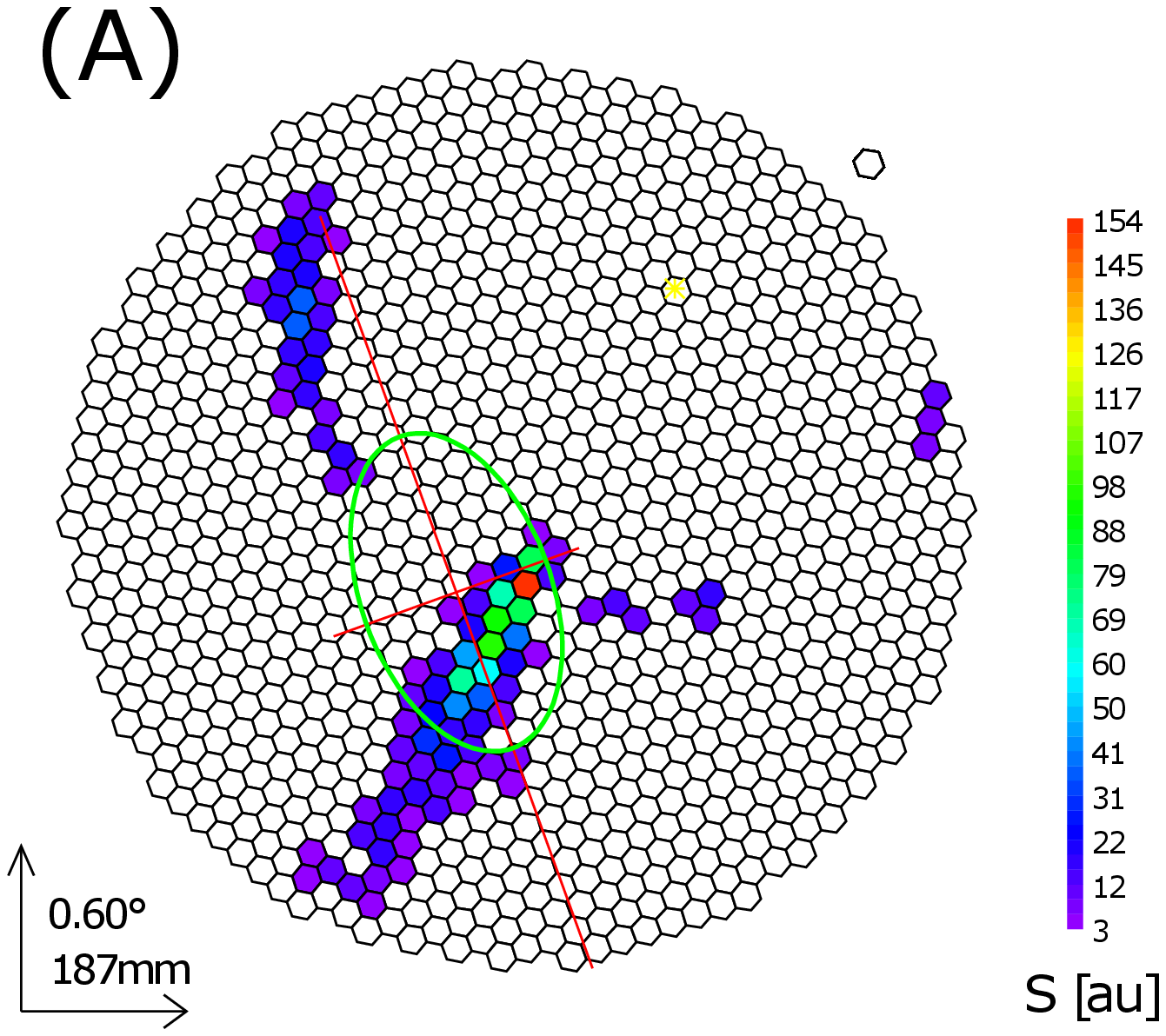}
\includegraphics [width=0.27\textwidth]{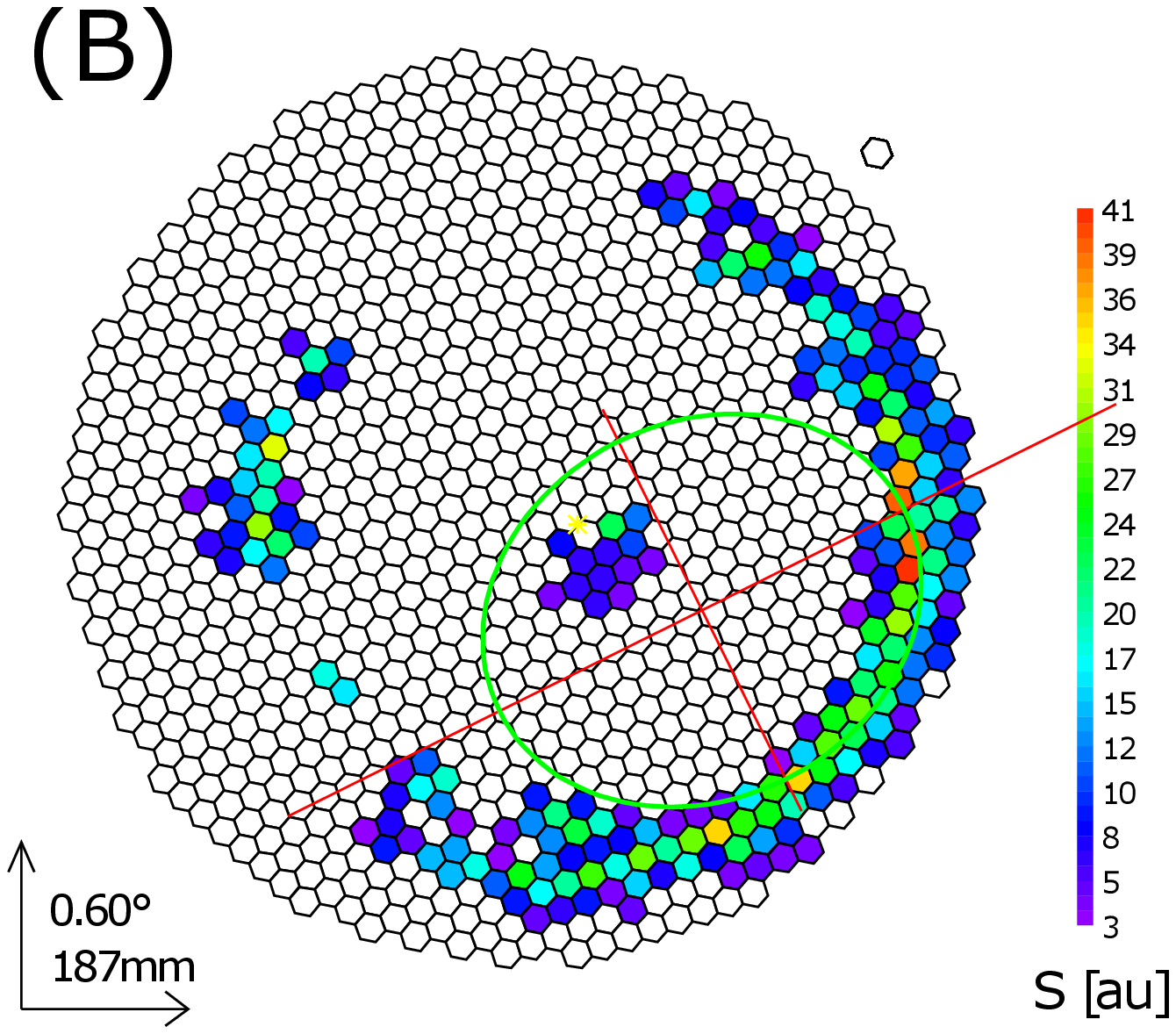}
\includegraphics [width=0.27\textwidth]{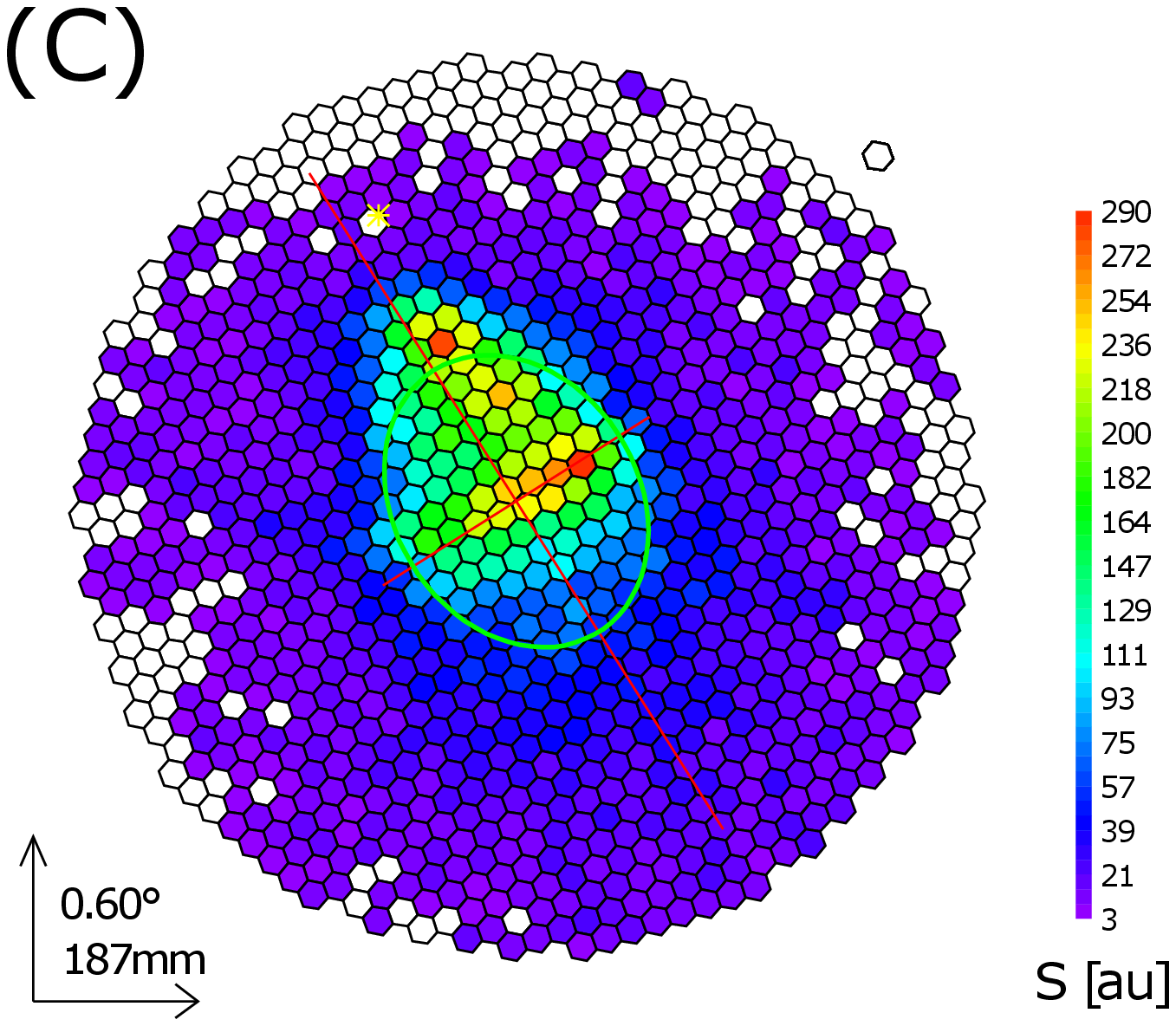}

\includegraphics [width=0.27\textwidth]{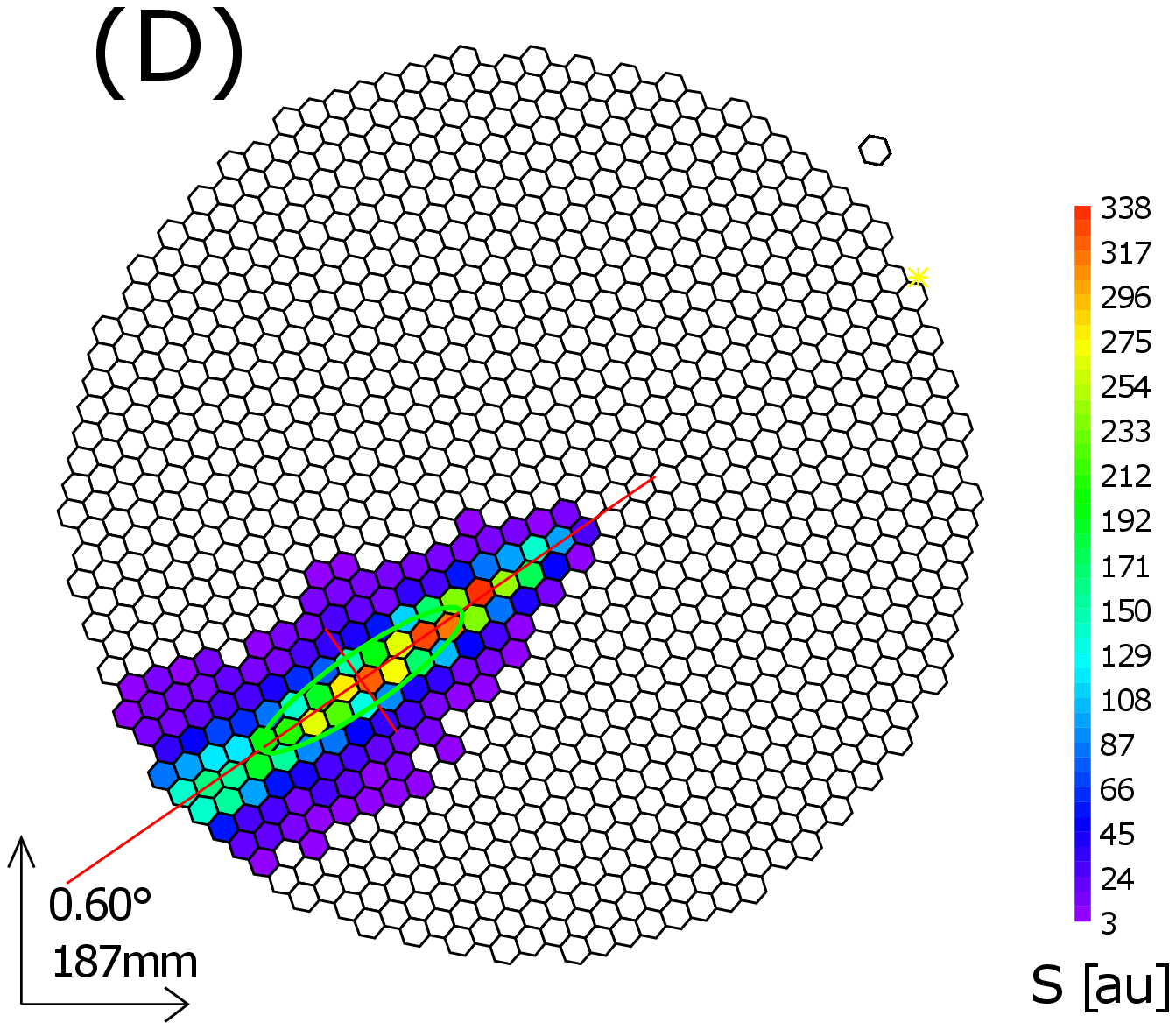}
\includegraphics [width=0.27\textwidth]{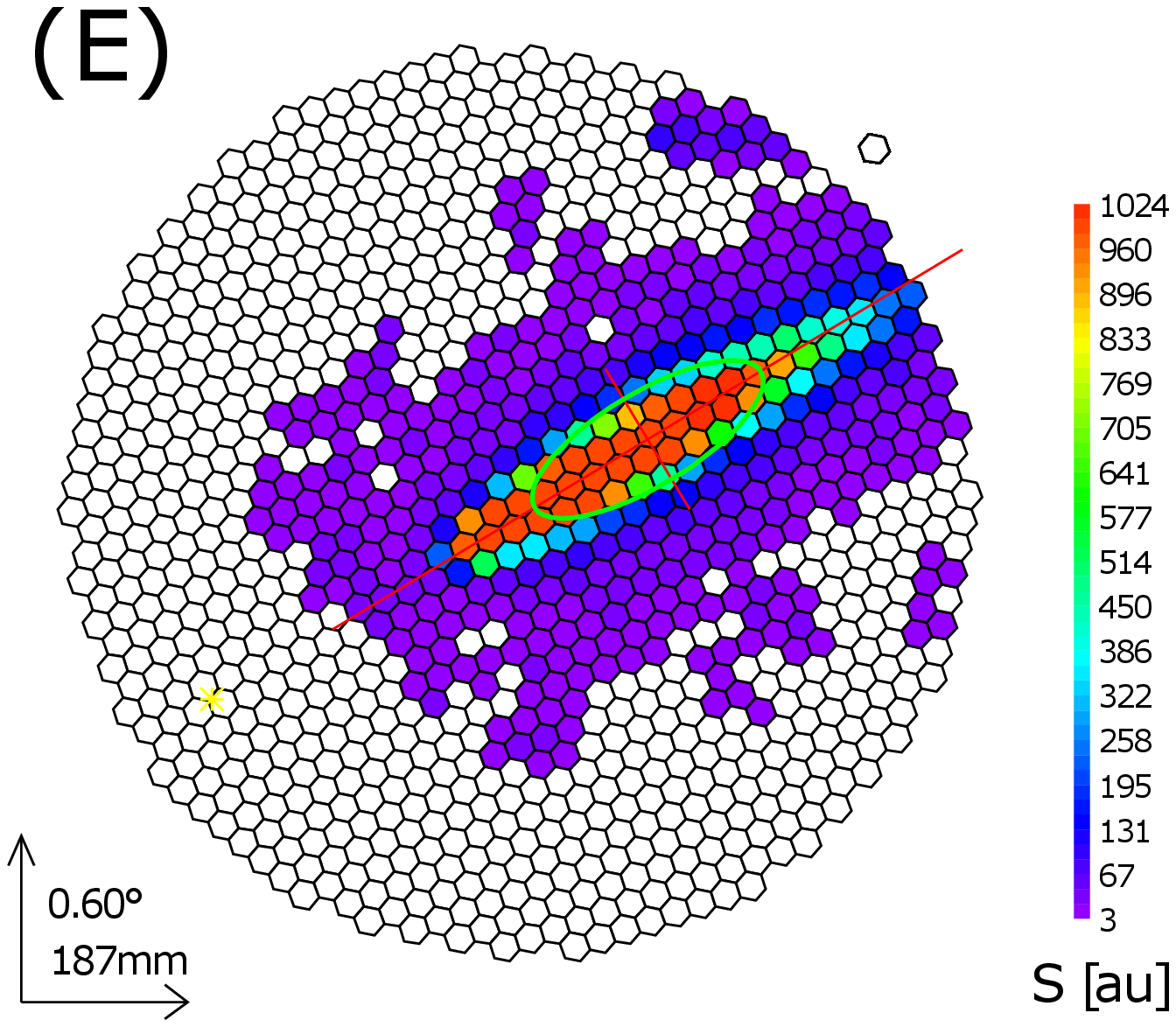}
\includegraphics [width=0.27\textwidth]{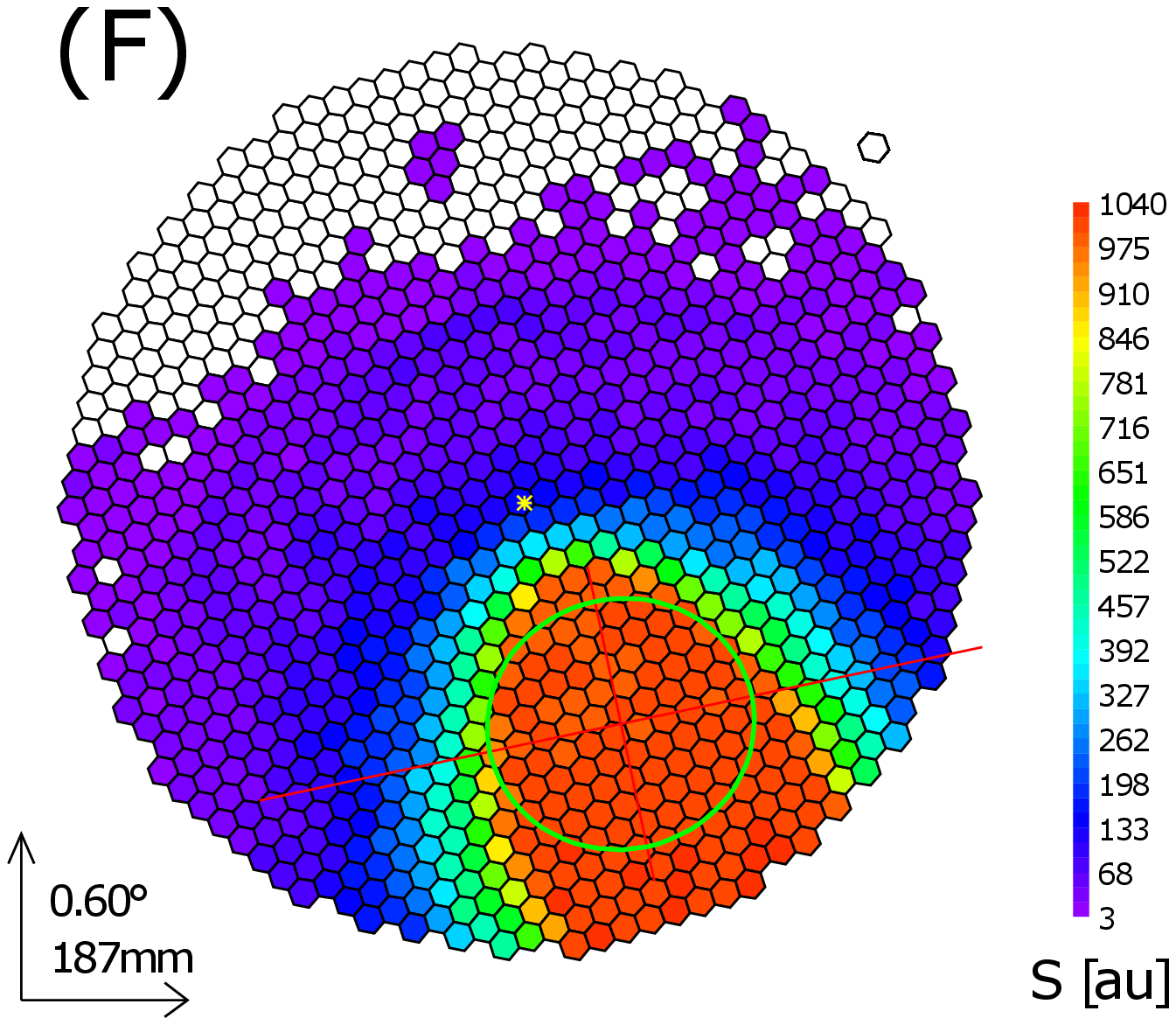}

 \caption{\small  Example of simulated shower images for 1 PeV protons (upper panels) injected at the top of the atmosphere (detector-to-proton distance  of about  800 km) and   1 PeV tau lepton (lower panels) decaying  close to the detector (about 50 kms)  for  zenith angle of $\theta\simeq 86^{\circ}$,
 as seen by one of  the MAGIC telescopes. 
(A) muon boundles producing a Cherenkov part of a ring and gamma-electron showers induced by the secondary muon decay in flight; (B) a ring from 
 a high energetic muon; (C) muons interacting  via radiative processes. For tau lepton  images  are shown  for  different tau decay channels: 
$\tau^{-} \rightarrow e^{-} \bar{\nu}_{e}\nu_{\tau}$ (D),  $\tau^{-} \rightarrow \pi^{-} \pi^{+} \pi^{-} \pi^{0} \nu_{\tau}$ (E), and   $\tau^{-} \rightarrow \pi^{-} \nu_{\tau}$ (F). }\label{fig::backimages}
\end{figure*} 

On the other hand, if flaring or disrupting point sources such as gamma ray bursts (GRBs) or active galactic nuclei (AGNs) are being pointed at, one can  expect neutrino fluxes producing
an observable number of events~\cite{Asaoka:2012em} and~\cite{upgoing_magic}. Also, for  sites with different orographic conditions, the acceptance for upward-going tau neutrinos is increased by the presence of mountains, which shield against cosmic rays and star light and serve as target for neutrino interaction leading to an enhancement in the flux of emerging tau leptons.

From the observational point of view, it is worth noting that the time that can be dedicated to this kind of observations  almost does not interfere 
with regular MAGIC gamma-ray observations of the sky, because the sea can be pointed even 
in the presence of optically thick clouds above the MAGIC site (high clouds). 
In fact, high-altitude clouds prevent the observation of gamma-ray sources
but still allow pointing the telescopes to the horizon. As an example, for the MAGIC site there
are up to about 100 hours per year where high clouds are present~\cite{frac}.  
\vspace{-0.30cm}
\section{MAGIC observations and Monte Carlo simulations}
 \vspace{-0.425cm}
Recently, the MAGIC telescopes has taken data  at very large zenith angles ($85^{\circ} <\theta < 95^{\circ}$) in  the direction of the sea (seeON), just above the sea (seeOFF) and towards  the Roque de los Muchachos mountain, as listed in  Table~1.  91\% of  the data were taken during nights characterized by optically thick high cumulus clouds, when normal $\gamma-$ray observations are usually worthless. Thus,  Table~1  also demonstrates,  that a large  amount of data ($\sim 39$ hrs) can be accumulated during such conditions, enhancing the overall duty cycle of MAGIC.
\begin{table}[t!]
\small
\center
\begin{tabular}{lccc}
\hline
 & seaOFF & seaON   & Roque     \\
\hline
Zenith angle $\theta$ ($^{\circ}$)&87.5 & 92.5$^{*}$ & 89.5  \\
Azimuth  $\phi$ ($^{\circ}$) &-30 & -30 & 170 \\
Observation time (hrs) & 9.2 & 31.5 & 7.5 \\
\hline
\end{tabular}
\caption{Summary of data taken at very large-zenith angles  by the MAGIC telescopes. 
$^*$ for this zenith  angle,  we expected the  maximal  acceptance for tau neutrinos~\cite{upgoing_magic}.}\label{tab:a}
\vspace{-0.425 cm}
\end{table} 
In order to study the signatures expected from neutrino-induced showers by MAGIC, a
full Monte Carlo (MC)  simulation chain was set up, which consists of three steps. First, the interaction of a given neutrino flux with the Earth and propagation of the resulting charged lepton through the Earth and the atmosphere is simulated using an extended version~\cite{goraanis} of the ANIS code~\cite{anis}. Second, the shower development of $\tau$-induced 
showers and its Cherenkov light production is simulated with CORSIKA~\cite{corsika}~\footnote{CORSIKA (version
6.99) was compiled with the CERENKOV, CURVED-EARTH, TAULEP, and THIN options, while the tau decay is simulated with PYTHIA.}. The results of the CORSIKA simulation are used as
inputs for the last step, i.e. the simulation of atmospheric extinction and the MAGIC detector response~\cite{mars}. It has to be said that,  we could not simulate  showers with  zenith angle $\theta>90^{\circ}$ when combining  CORSIKA  and MAGIC simulations due to unsolved technical subtleties with the MAGIC simulations. Therefore, here we have used   simulations for zenith angles in the range between $ 86^{\circ}$  and $90^{\circ}$ to study properties of upward-going tau neutrino showers. This is a reasonable assumption, because  the response of IACTs to Cherenkov light from showers of a same energy and equal column depth only slightly depend on the zenith angle, as has been already demonstrated in~\cite{gora:2016}.

\begin{figure*}[t!]{}
\begin{minipage}[b]{0.49\textwidth}
\includegraphics [width=0.95\textwidth,height=4.8cm]{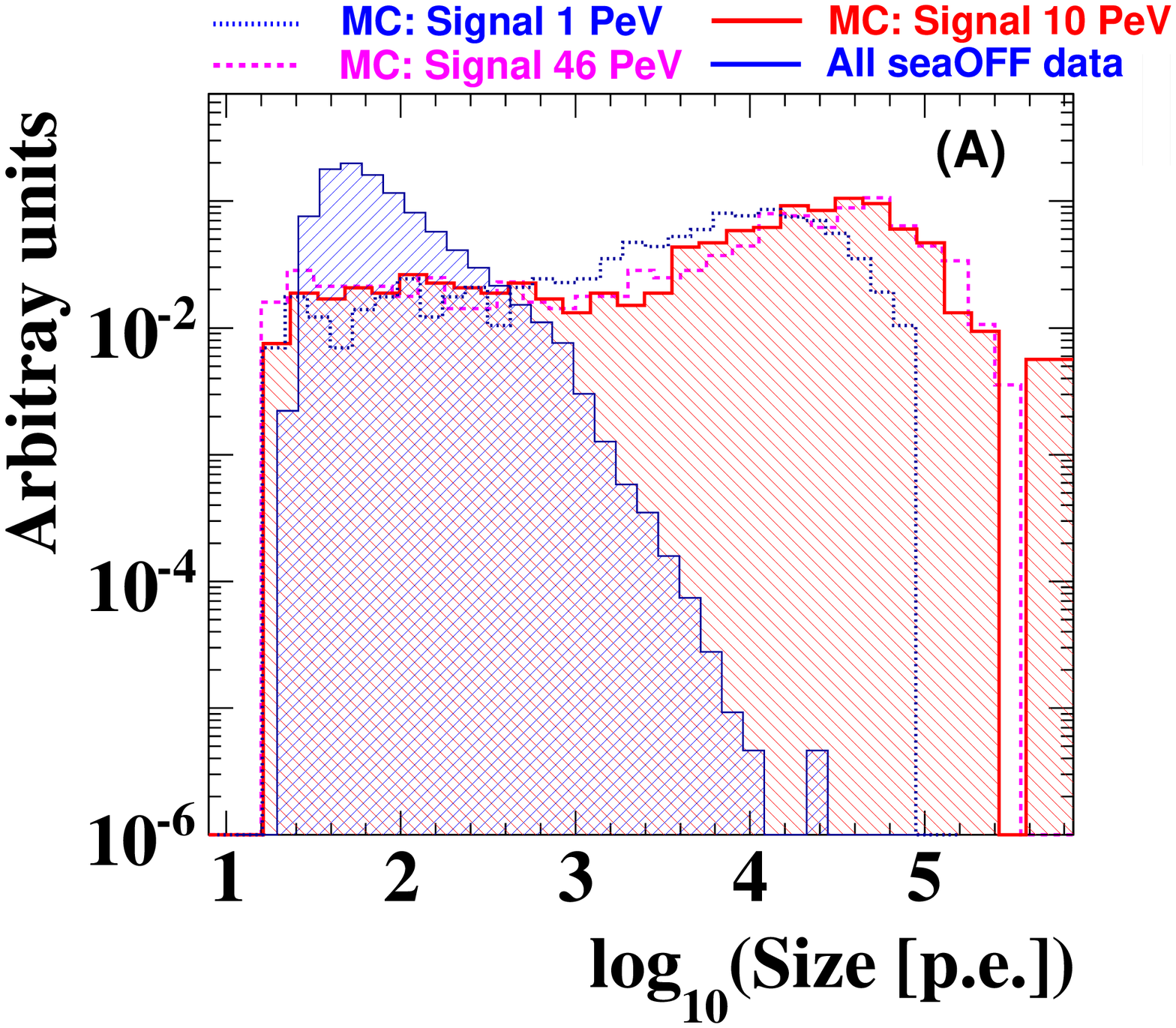}
\includegraphics [width=0.95\textwidth,height=4.8cm]{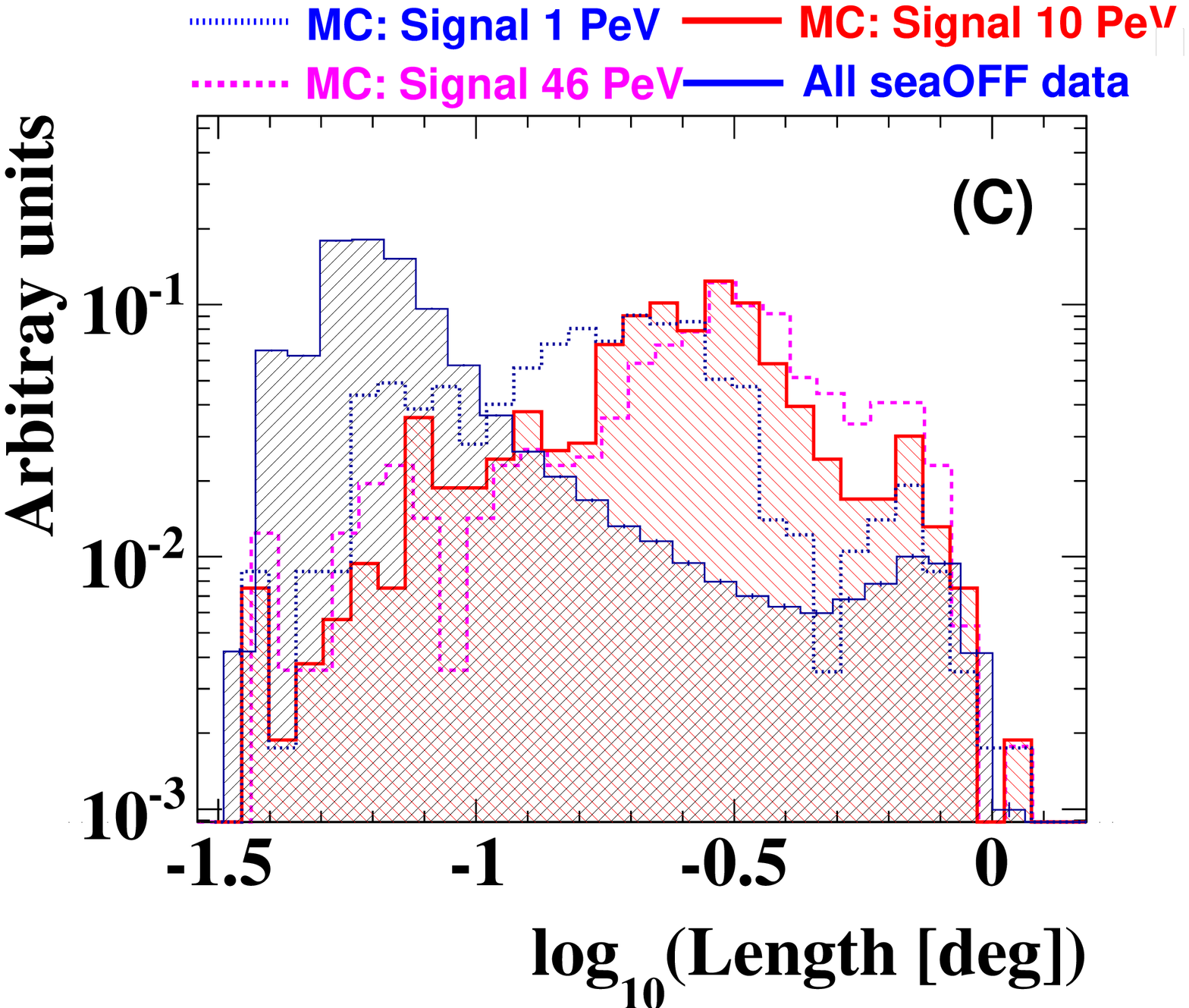}
 \end{minipage}
\begin{minipage}[b]{0.49\textwidth}
\includegraphics [width=0.95\textwidth,height=4.8cm]{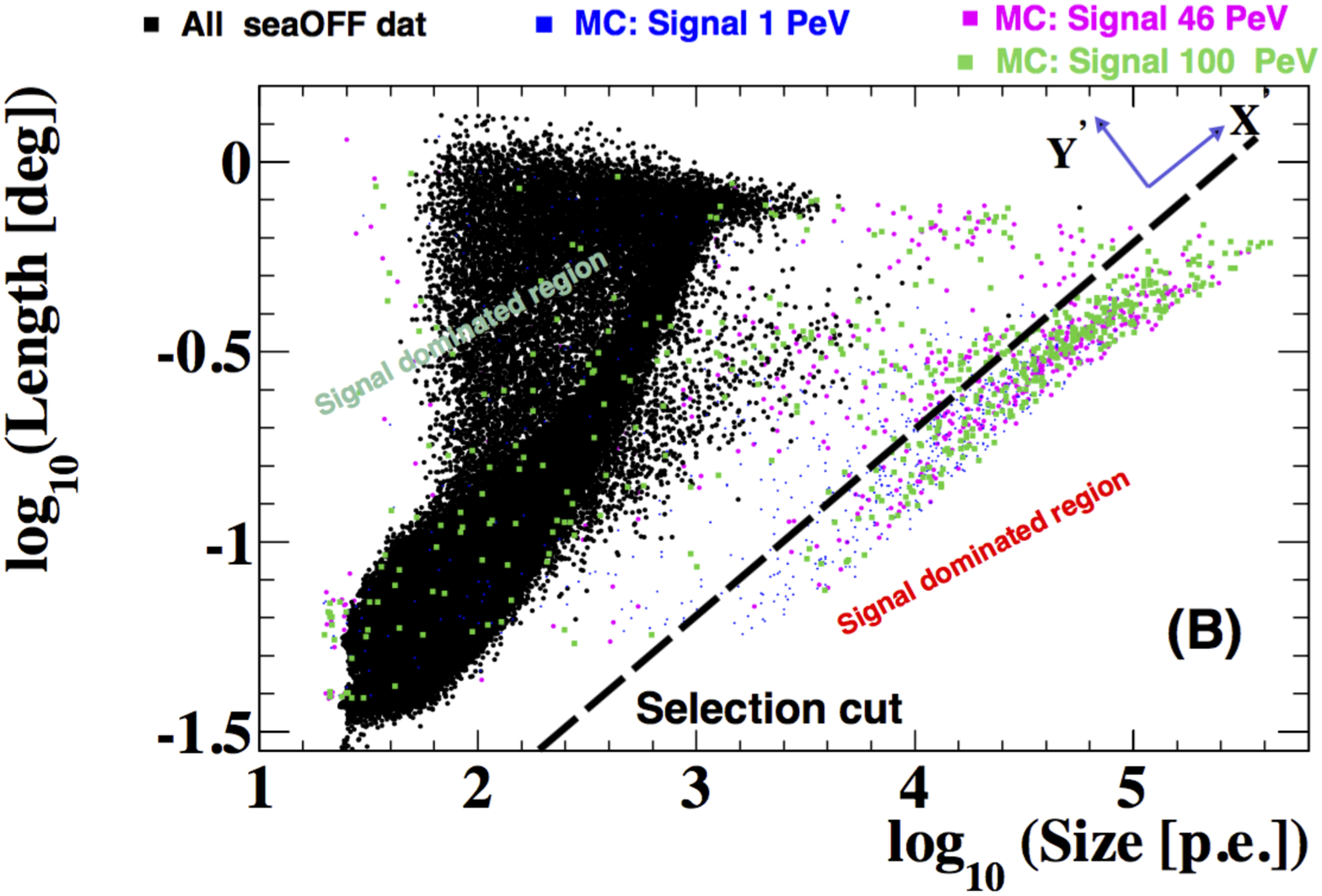}
\includegraphics [width=0.95\textwidth,height=4.8cm]{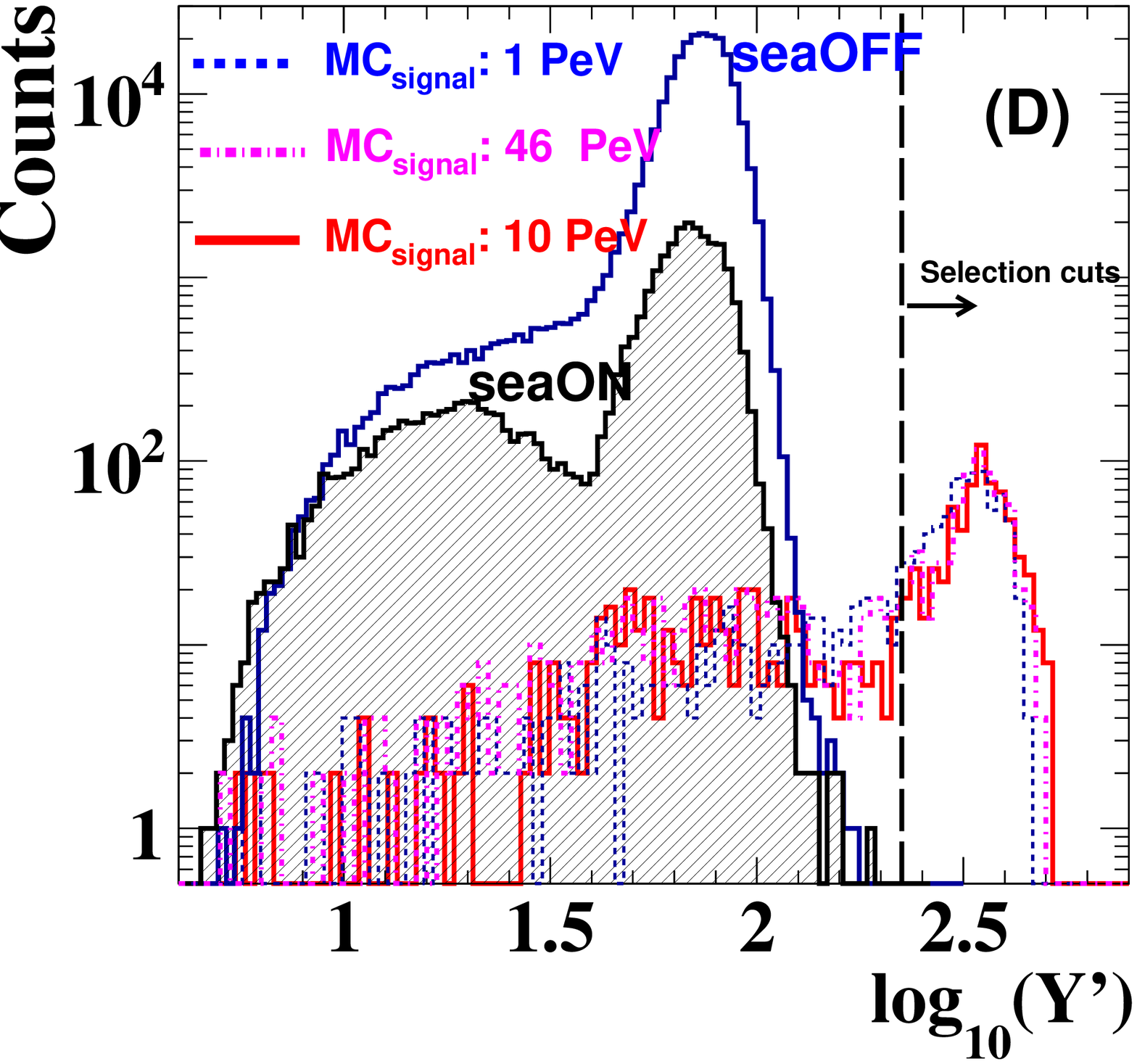}
\end{minipage} 
\caption{ \small (A),(C): normalized distribution of Hillas parameters for deep $\tau$-induced showers 
 with zenith angle $\theta=87^{\circ}$ and impact distances less than 0.3 km (signal) and data taken toward the seaOFF  direction (about  $2.2\cdot 10^{5}$ background  events). (B): scatter plot of the Hillas $Length$ parameter  as a function of the Hillas $Size$ parameter. Dashed line indicates the selection cut used in this work to calculate the identification efficiency for deep $\tau$-induced showers. Note that the region below the discrimination cut is characterized by the absence of background events. (C): The one-dimensional distribution of seaON, seaOFF and signal MC obtained in the direction perpendicular to the selection cut. The new coordinate  for data were obtained from  the following formula: $\log_{10}(Y{'})=\log_{10}(Size[p.e.])*\cos(\alpha)-\log_{10}(Length[deg])*\sin(\alpha)$, where $\alpha=63.435^{\circ}$.  Note that  above the selection cut ($\log_{10}(Y^{'})>2.35$)  zero  neutrino candidates are found. For showers with the  larger impact distances (0.3-1.3 km) a slightly  relaxed cut was used:  $\log_{10}(Y^{'})>2.10$.} \label{fig::dist}
\vspace{-0.5cm}
\end{figure*}
In order to compare simulated  images on the MAGIC camera plane with  those obtained from data, we also simulated inclined showers induced by protons. Proton showers can mimic the Cosmic  Rays (CRs) background for Cherenkov telescopes. In case of showers induced at the top of the atmosphere
 and observed at large zenith angles,  the hadronic and electromagnetic component of extensive air showers (EAS) are almost  absorbed because  of the deep horizontal column depth ranging from $\simeq 10^{4}$   to $5\times10^{4}$ g cm$^{-2}$ at such directions.  Thus,  for proton induced showers with lower energies (tens of GeVs) only a  few pixels will be triggered by the camera yielding  
dimmer and smaller images.  Only for  higher energy CRs (above $1 \times 10^{15}$ eV), high energetic  muons (tens to hundreds of GeV)  from the first stages  of the shower development in the atmosphere  or   muon bundles   from later stages can reach the detector and  produce larger  shower images, see Figure~\ref{fig::backimages} (A).  On the contrary,   for  high energetic  muons  the  shower image  in the camera  will   mostly contain a single  muon ring,  if the muon hits the telescope mirror, or  an incomplete  ring (arc), see Figure~\ref{fig::backimages} (B). 
At larger zenith angles, for   highest energetic muons ( > 1 TeVs)   the interaction length  due 
to  $e^+e^-$-pair production, bremsstrahlung  and photonuclear scattering  is comparable to the depth of the  atmosphere, thus
we also expect a few shower events per night, coming from  interacting  muons  via  radiative processes~\cite{kiraly,kiraly2}. As our simulation shows, if  one of these sub-showers  from radiative  interactions are induced close to a detector, this can  lead to  a strong
flash of  Cherenkov light and  a  bright  image on the camera  (see  Figure~\ref{fig::backimages} (C) as an example).  
 All  classes of simulated events discussed above   were also  identified  in the  MAGIC data taken  at  very large zenith angles. 
 
 In case of tau  leptons,  the expected signature on the camera  depends on the tau decay channel. The tau lepton decays  
mostly into  electrons, muons or charged and neutral pions~\cite{taudeacy}. As an example, in Figure~\ref{fig::backimages} (D)-(F)  we show  
simulated shower images  for the  1 PeV tau lepton decaying into an electron, pion (or several pions)  close to the detector. In general, for  such a geometry,  the shower images on the camera   have a much larger  size  and contain  many more photons  compared to the  proton images. The tau lepton can  also decay into  muons.
However, at high energies (> 1 PeV) the moun  has  a large interaction length, more than  a few  thousand kilometers in air, thus it  mostly
interacts with the atmosphere through secondary bremsstrahlung processes, which blur the muon image and make the ring only poorly visible.
\vspace{-0.370 cm}
\section{Discrimination of $\tau$-induced showers }
 \vspace{-0.405 cm}
Each simulated event recorded and calibrated consists of a number of photoelectrons (p.e.) per camera pixel. The cleaned camera image  is characterized by a set of image parameters introduced by M. Hillas in~\cite{hillas}. These parameters provide a geometrical description of the images of showers and are used to infer the energy of the primary particle, its arrival direction, and to distinguish between $\gamma-$ray and hadron induced showers. In the following we  study  these parameters also  in the case of  deep $\tau$-induced simulated showers and compare the corresponding distributions with the data. 

 As previously mentioned  the MAGIC telescopes took    data also at zenith angle 87.5$^{\circ}$ (seeOFF). For this zenith angle we expect  a negligible signal (neutrino events) contribution  compared  to the sea. Thus seeOFF data  were  used  to estimate 
the  background  and to construct the selection criterion   to indentify  tau-neutrino showers. In addition,
the  rate  of stereo  seeOFF events is about 27 times  larger   ($\sim$ 4.6 Hz) than  for seaON ($\sim$0.17 Hz) observations. Thus these observations provide 
high-statistics background estimates for  about  30 hours of seaON data.

In  Figure~\ref{fig::dist} (A),(C) the distribution of the Hillas parameter {\it Size} and {\it Length} 	 for deep $\tau$-induced simulated showers  are shown, in comparison to seeOFF data. In general, these parameters  depend on the  geometrical distance of the shower maximum to the detector, which for deep $\tau$-induced showers is much smaller (a few tens of kilometers) than for showers from CRs  interacting  at  the top of the atmosphere (a few  hundred kilometers).  This geometrical effect leads to a very  good separation of close ($\tau$-induced) and far-away (data) events in the Hillas parameter phase space.  
 
Figure~\ref{fig::dist} (B) shows the scatter plot of the {\it Length} parameter as a function of the {\it Size} parameter  for our MC signal simulations and  seeOFF data. We can see that   with these two parameters only, we can  easily identify  a region with no background events. This plot shows that MAGIC  can discriminate deep $\tau$-induced  showers from the background of high zenith angles hadronic showers. In   Figure~\ref{fig::dist} (D),  we show the one-dimensional distribution of seaOFF, seeON and MC signal simulation  projected onto the  line perpendicular to our selection cut.  As it is  seen  we did not find  any neutrino candidate, if the selection cut is applied  to  all  seeON data. 

It is  also worth noting that the shape of the Hillas distributions for $Size$ and $Length$ agree reasonably well with our MC background simulation, i.e. with simulated proton induced showers. In addition the shape of these   distributions  for seaOFF data  and
seaON data  is  very similar,  indicating  a universal  behaviour  of Hillas distributions at these zenith angles. This universality
 also confirms our previous assumption when performing  MC simulations  only at large zenith angles,  in order to study the response of MAGIC  to upward-going $\tau$-induced showers.
\vspace{-0.325cm}
\section{Aperture and event rate calculations }
\begin{figure*}[t!]
 \centering
 \includegraphics [width=0.49\textwidth,height=4.8cm]{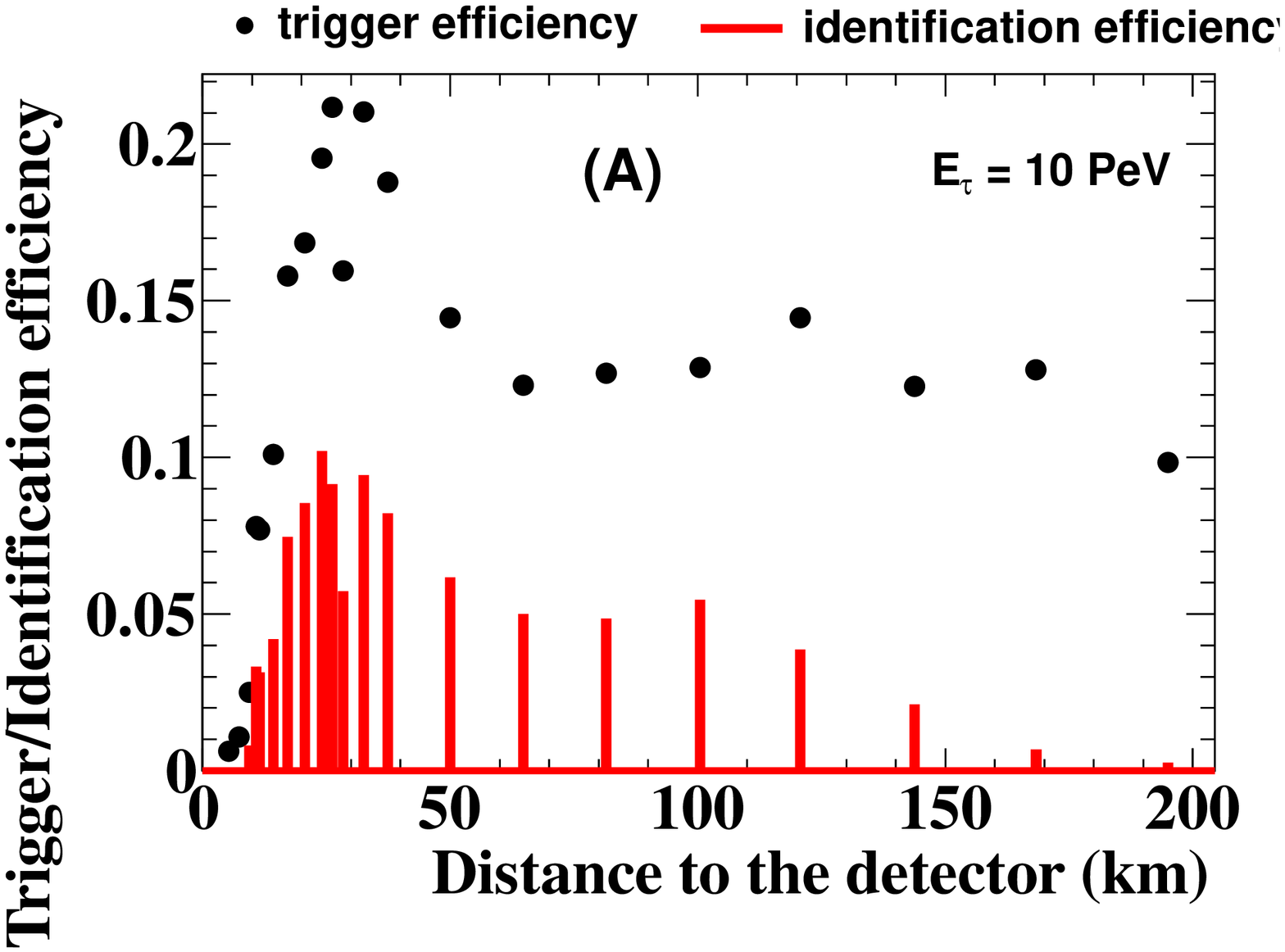}
  \includegraphics [width=0.48\textwidth,height=4.8cm]{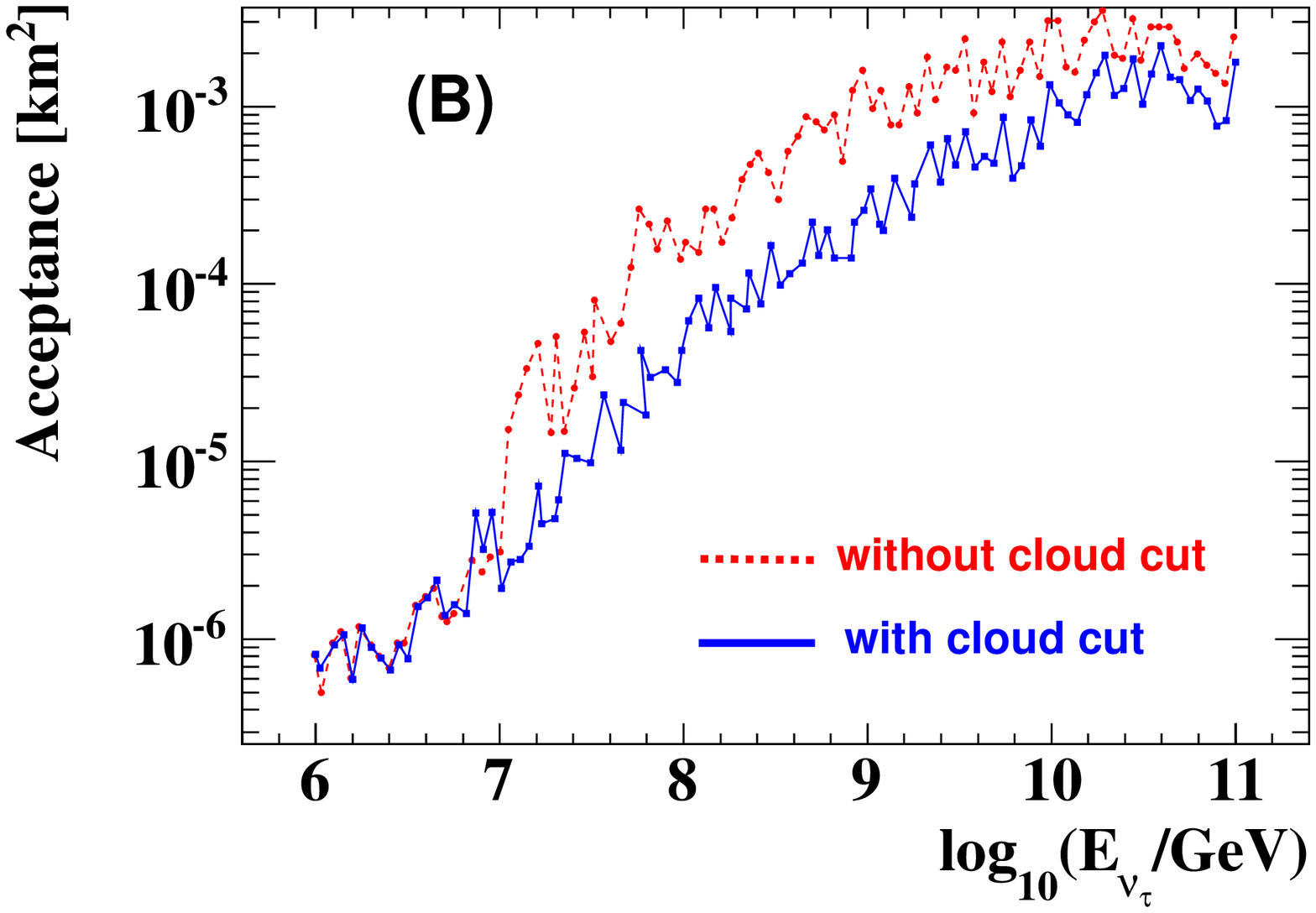}
  \vspace{-0.25cm}
 \caption{\small Left panel: trigger/identification efficiency for MAGIC as a function of the distance to the telescope. This is an average over  simulated   showers 
 with impact distance smaller than 1.3 km and for zenith angle $86^{\circ}$ . Right panel: MC acceptance for point sources, $A^{\mathrm{PS}}(E_{\nu_\tau})$, to  earth-skimming tau neutrinos as estimated for the MAGIC site. For MAGIC pointing at $\theta=92.5^{\circ}$ and  $\phi=-30^{\circ}$, the local orographic condition are included. }\label{fig1}
\vspace{-0.55cm}
\end{figure*}

\vspace{-0.325cm}
The total observable event rates (number of expected events) were calculated as 
$N=\Delta T \times \int_{E_{\mathrm{th}}}^{E_{\mathrm{max}}} A^{\mathrm{PS}}(E_{\nu_\tau})\times\Phi(E_{\nu_\tau})\times
dE_{\nu_\tau}$ where  $\Delta T$  is the observation time,
and $A^{\mathrm{PS}}(E_{\nu_\tau})$ the  point source acceptance and $\Phi(E_{\nu_\tau})$ the expected neutrino flux.
 The detector  acceptance for an initial neutrino energy $E_{\nu_\tau}$ is calculated as:
$A^{\mathrm{PS}}(E_{\nu_\tau}, \theta,\phi)  =N_{\mathrm{gen}}^{-1} \times \sum_{i=1}^{N_{FOV cut}}
   P_{i}(E_{\nu_\tau},E_{\tau},\theta) \nonumber  
    \times  A_i(\theta) \times T_{\mathrm{eff},i}(E_{\tau},r,d,\theta)$
where $\theta$, $\phi$ are the simulated zenith and azimuth pointing angles of the MAGIC telescope,  $N_{\mathrm{gen}}$ is the number of generated neutrino events. $N_{\mathrm{FOV cut}}$ is the number of $\tau$ leptons with energies $E_{\tau}$ larger than the threshold energy $E_{\mathrm{th}}=1$\,PeV and  with 
 estimated  position of the shower maximum in the FOV of the MAGIC  telescope. In addition the impact distance of showers should be smaller than 1.3 km. $P(E_{\nu_\tau},E_{\tau},\theta)$ is the probability that a neutrino with energy $E_{\nu_\tau}$ and zenith angle $\theta$  produces a lepton with energy $E_{\tau}$, this probability was used as a "weight" of the event. $A_i(\theta)$ is the physical cross-section area of the interaction volume seen by the neutrino, simulated by a cylinder with radius of 50 km and height 10 km. $T_{\mathrm{eff,i}}(E_{\tau},r,d,\theta)$ is the trigger and identification efficiency for $\tau$-lepton induced showers with the decay vertex position at  distance $r$ to the telescope  and  impact distance  $d$. The trigger efficiency in an energy range interval $\Delta E$, is defined  as the  number of simulated showers with positive trigger decision and fulfilling our selection criterion shown in  Figure~\ref{fig::dist} (D) over the total number of generated showers for  fixed zenith angle $\theta$, initial energy of primary particle $E_{\tau}$, and the impact distance range~\footnote{The  impact  distance of simulated showers  was randomized in  CORSIKA simulations  (by using CSCAT option)  and later  Cherenkov telescope orientation  for such showers 
was randomized  over the MAGIC camera FOV.}.  As an example  Figure~\ref{fig1} (A) gives the trigger/identification efficiency as a function of  the distance to the detector.  For smaller distances ($d < 20$ km)  the trigger/identification  efficiency drops due to the fact that the shower maximum is too close to the detector or the shower does not reach yet the maximum of shower development,  decreasing  the amount of Cherenkov light seen by the telescopes. In general, the plot   provides  an estimate  of the typical distance  for $\tau$-induced showers seen by MAGIC. 
In Figure~\ref{fig1} (B)  we show  an estimate  for the MC point-source acceptance  to tau neutrinos. The simulation  of the aperture
includes  the density profile of the Earth and a 3 km deep water layer around the La Palma island. The water layer is important because it leads 
to about  a factor two (in the energy range 10-100 PeV) smaller acceptance than for the spherical Earth calculations with the rock density of about 2.65 g/cm$^2$.  In Figure ~\ref{fig1} (B)   we also show the acceptance, when a cloud  layer is  included in our simulation i.e. the quasi-stable sea of cumulus between 1500 and 1900 m a.s.l.,  usually present  at MAGIC site due to the  temperature inversion \footnote{To estimate the effect of this we assumed that all decaying tau leptons  below 1500 a.s.l. are discarded in the  acceptance  calculations. Thus we 
assume that  for such induced shower  all Cherenkov light is absorbed  when it passes the range from 1.5-1.9 km a.s.l. }. As we can see from the plot 
 this cloud cut  leads  also to a smaller (about factor two) acceptance. 

As we already mentioned, Cherenkov telescopes can be sensitive to tau neutrinos from fast transient objects like GRBs  or AGNs.
 In  other words flaring sources, including GRBs, Tidal Disruption Events or the  so-called  Low Luminosity GRB (LLGRBS),
 can provide a boosted flux of neutrinos.   In Table~\ref{table4} the expected event rates for MAGIC    for  fluxes from AGN benchmark models, shown in \cite{gora:2016}, are listed.  The rate is calculated for tau neutrinos   assuming that the source is in  the MAGIC  telescope FOV  for a period of 3 hours.  For   Flux-3 and Flux-4  (i.e. those models covering the energy range beyond   $\sim 1\times10^{8}$ GeV) the event rate  is   at the  level of $4 \times 10^{-5}$. For neutrino fluxes covering the energy range below $\sim 5\times10^{7}$\,GeV (Flux-1, Flux-2, Flux-5), the number of expected events  is  not larger than  $1 \times 10^{-5}$.

From the estimated acceptance with cloud cut, the sensitivity for an injected spectrum $K\times\Phi(E_{\nu})$ with a known shape $\Phi(E_{\nu})$ was                                                                                                      
calculated. The 90\% C.L. on the value of $K$, according to ~\cite{limit} is $K_{90\%}=2.44/N_\mathrm{Events}$, with the assumption of                                                                                                                   
negligible background, zero neutrino events being observed by the MAGIC  during  sea observations, and  in case of the flux $\Phi(E_{\nu})=1 \times 10^{-8} E^{-2} \mbox{ [GeV cm$^{-2}$ s$^{-1}$]} $,  the sensitivity for a point source search  is $E_{\nu_\tau}^{2}\Phi^{PS}(E_{\nu_\tau}) < 1.7   \times 10^{-4}$ \mbox{ [GeV cm$^{-2}$ s$^{-1}$]} in the range from 2 to 1000 PeV. The  sensitivity is calculated  for  the  expected number of tau neutrino events equal to  $ N_\mathrm{Events}=1.4 \times10^{-4}$, based on the result listed in  Table~\ref{table4} for Flux-5, and   for  30 hours of observation time. The  sensitivity  can be improved   about  two orders of magnitude   for  larger observation time ($\sim 300$ hrs) and in case of  a strong flare with a neutrino flux given by  Flux-4, reaching  the value  $E_{\nu_\tau}^{2}\Phi^{PS}(E_{\nu_\tau}) < 5.8\times 10^{-6}$ \mbox{ [GeV cm$^{-2}$ s$^{-1}$]}  i.e. the level of the down-going analysis  of the Pierre Auger Observatory ~\cite{auger}.
\vspace{-0.25cm}
\section{Summary and conclusion} 
\vspace{-0.225cm}
 Taking into account our sensitivity estimate, the observational program for tau neutrino searches seems to be challenging, but in principle not impossible to pursue. We would like to stress that  observation time can be accumulated  during periods with high clouds,  when those instruments are not used for gamma-ray observations. Note also, that  this kind of search, as shown in this proceeding, is  basically background free,  so the tau neutrino sensitivity increases linearly with  the observation time. Finally, the  next-generation Cherenkov telescopes, i.e. the Cherenkov Telescope Array, will exploit its much larger FOV (in extended observation mode), and much larger effective areas.
 \begin{table*}[bt!]
\small
\caption{\small Expected event rates for the MAGIC detector   in case of AGN flares. Flux-1 and Flux-2 are predictions for neutrinos from $\gamma$-ray flares of 3C 279~\cite{2009IJMPD}. Flux-3 and Flux-4 are  predictions for PKS~2155-304 in low-state and high-state, respectively~\cite{Becker2011269}. Flux-5 corresponds to a prediction for 3C~279 calculated
in~\cite{PhysRevLett.87.221102} and  is  at  a similar level in the PeV energy range like  the flux reported by IceCube for
astrophysical high-energies  neutrinos~\cite{lasticecube}.}\label{table4}
\center
\vspace{-0.4cm}
\begin{tabular}{ccccccccc}
\hline
 &\bf Flux-1  &\bf Flux-2& \bf  Flux-3 & \bf Flux-4 & \bf Flux-5 \\
 & \bf ($\times 10^{-5}/3$ hrs)   & \bf ($\times 10^{-5}/3$ hrs)  &  \bf ($\times 10^{-5}/3$ hrs)    &\bf ($\times 10^{-5}/3$ hrs)  &\bf ($\times 10^{-5}/3$ hrs)   \\
\hline
$N_{\mathrm{Events}}$  &1.3 & 0.7 & 0.42 &4.2 &1.4\\
\hline
\vspace{-0.4cm}
\end{tabular}
\vspace{-0.45cm}
\end{table*}

\end{document}